\documentstyle[preprint,aps]{revtex}
\begin{document}
\draft
\preprint{IUCM95-001}
\title{Interactions, Localization, and the Integer Quantum Hall Effect}

\author{S.-R. Eric Yang$^1$ and A.H. MacDonald$^2$}
\address{$^1$Institute for Microstructural Sciences, National Research
Council of Canada, Ottawa, Ontario, Canada, K1A 0R6\\
$^2$Physics Department, Indiana University, Bloomington, IN 47405, USA \\}
\date{\today}
\maketitle

\begin{abstract}

We report on numerical studies of the influence of Coulomb interactions on
localization of electronic wavefunctions in a strong magnetic
field.  Interactions are treated in the Hartree-Fock approximation.
Localization properties are studied both by evaluating participation ratios
of Hartree-Fock eigenfunctions and by studying the boundary-condition
dependence of Hartree-Fock eigenvalues.  We find that localization properties
are independent of interactions.  Typical energy
level spacings near the Fermi level and the sensitivity of those
energy levels to boundary condition show similar large
enhancements so that the Thouless numbers of the Hartree-Fock
eigenvalues are similar to those of non-interacting electrons.

\end{abstract}

\narrowtext

In studies of the integer quantum Hall effect it is usually assumed
without justification
that Coulomb interactions between electrons can safely be ignored.
At strong magnetic fields, eigenfunctions of the single-particle
Hamiltonian are localized by a disorder potential
at almost all energies and the localization length diverges
as $|E - E_{cN}|^{-7/3}$ at a discrete set of critical energies
$E_{cN}$ near the center of each disorder-broadened Landau level\cite{lit20}.
The critical energies are similar to the mobility edges
of the Anderson metal-insulator transition, except that the localization length
is finite both for $E > E_{cN}$ and for $E < E_{cN}$.
Theory suggests that at $T=0$ the Hall conductivity ($\sigma_{xy}$)
jumps by $e^2/h$ each time the Fermi energy $E_F$ crosses
one of the critical energies, that the dissipative conductivity ($\sigma_{xx}$)
is zero unless $E_F=E_{cN}$, and\cite{lit60,lit50,lit30,gammel}
that $\sigma_{xx}=\sigma_{xy}=\frac{e^2}{2h}\equiv\sigma_c$ when
$E_F=E_{cN}$.  All these suggestions are supported
by experimental\cite{lit80} studies, except possibly\cite{possib} for the
predicted
universal value of the critcal dissipative conductivity $\sigma_{xx}$.
In this Letter, we report on numerical studies of the
integer quantum Hall effect which include Coulomb interactions
at the lowest consistent level by performing Hartree-Fock
calculations in the presence of random disorder potentials.
The possible importance of Coulomb interactions is suggested
by studies in three-dimensions
at zero magnetic field where the metal-insulator transition
can be profoundly altered by interactions\cite{lit40}, and by the
presence of a Coulomb
gap in the integer quantum Hall regime\cite{lit45,lit120}.
However, general scaling considerations suggest\cite{lit50,lit60} that for the
quantum Hall effect the critical conductivities are independent
of Landau level index $N$, details of the disorder potential, and
of interactions.

The concept of a localization length can be generalized to the
case of interacting electrons by defining it in terms of the
dependence of the imaginary part of the the one-particle Greens function
on spatial coordinates.  In the Hartree-Fock approximation,
this change is accomplished by replacing the single-particle eigenfunctions
in a random potential by the self-consistent Hartree-Fock
eigenfunctions.  We have studied the localization properties of these
eigenfunctions by examining their participation ratios and find
no qualitative changes.  This result is  reasonable since
the self-consistent Hartree-Fock Hamiltonian is always equivalent
to the single-particle Hamiltonian for a non-interacting system with
an effective random potential which however presumably has smaller long
wavelength
components because of screening effects.  (The exchange potential is
non-local but its effect is equivalent\cite{mgdm} to that of a local
potential.)
Universality of localization
properties in the absence of interactions would then seem to imply no
change in critical localization properties in the Hartree-Fock
approximation.

The influence of interactions on the critical
conductivity is difficult to determine, even within the Hartree-Fock
approximation.  The numerical evaluation of $\sigma_{xx}$ necessitates
the inclusion of interaction vertex corrections by summing ladder
diagrams\cite{hfvertex} in the presence of disorder.  We have
avoided this considerable obstacle by appealing to the close relationship
between $\sigma_{xx}$ and the more easily calculated
Thouless numbers which are defined in terms of the dependence
of eigenenergies on the boundary conditions of the system.
The definition we use for the Thouless number is:
\begin{equation}
g_T(E) \equiv \frac{\langle \delta E \rangle}{\Delta E }
\label{eq:thouless}
\end{equation}
where $1/\Delta E = L^2 D(E) $ is the average level spacing calculated from
the density of states per unit area $D(E)$,
$\langle \delta E \rangle $ is the geometric mean of
the eigenvalue differences between periodic and antiperiodic boundary
conditions
and $L^2$ is the area of the finite system.
For non-interacting electrons the Kubo formula for the conductivity
can be written as
\begin{equation}
\sigma_{xx} = \pi e^2 L^2 \hbar D^2(E_F) \langle |v_x|^2 \rangle
\label{eq:mott}
\end{equation}
where $\langle |v_x|^2 \rangle$ is an average\cite{lit90} over
matrix elements of the velocity operator between states at energies
near $E_F$.  It follows from simplifying assumptions about
correlations between velocity matrix elements and energy level spacings
that
\begin{equation}
\sigma_{xx} = \frac{e^2}{h}\frac{\pi g_T(E)}{2}
\label{eq:sigmaxx}
\end{equation}
both at\cite{lit90} $B=0$ and at\cite{ando,hanna} strong fields.  (In the
metallic limit a rigorous relationship between $\sigma_{xx}$ and a suitably
defined Thouless number can be established\cite{lit100} at $B=0$ for
non-interacting electrons.)
Our numerical calculation is motivated by the expectation that within
the Hartree-Fock approximation a
qualitative relationship between $g_T(E_F)$ and $\sigma_{xx}$ will
survive interactions.  Hartree-Fock eigenvalues change with
boundary conditions\cite{lit90,ando}
directly because of the velocity matrix elements between Hartree-Fock
eigenfunctions and indirectly because of the change in the exchange potential
produced by the changed eigenfunctions.  The former effect is
related to the `bubble' diagram for the conductivity just as it is
without interactions, while the latter effect is related to
vertex corrections in the Hartree-Fock approximation for the conductivity.
In the integer quantum Hall regime $D(E_F)$ is strongly suppressed
by interactions\cite{lit45,lit120} just as it is for disordered electrons
at $B=0$.
In view of Eq.(~\ref{eq:mott}), this suggests that $\sigma_{c}$ could
be different in interacting electron systems.  We have found that
a remarkable cancellation occurs in which the change in
$g_T(E_F)$ due to the decrease in $D(E_F)$ (increase in $\Delta E$)
is cancelled by increased sensitivity of the Hartree-Fock eigenvalues
to boundary conditions.  To the precision of our calculations,
$g_T(E_F)$ is unchanged by interactions.

In the model used for our numerical calculations $N$ electrons in the lowest
Landau level are confined to a square of area
$L^2 = 2\pi \ell^2 N_\varphi = N_\varphi \Phi_0/ B$. $N_\varphi$ is the number
of single-particle states in the lowest Landau level, $\Phi_0 \equiv
h c /e $ is the magnetic flux quantum and the
magnetic length, $\ell$ is defined by these relations.
Our model disorder consists of a randomly located delta-function
`impurities' with a random strength uniformly distributed between
$- \lambda$ and $\lambda$.   For this model\cite{ando} the energy scale which
characterises the Landau level width is
$\Gamma = (\lambda^2N_I /l^2L^2)^{1/2}$ where $N_I$ is the number
of impurities.
The relative strength of Coulomb interactions and
disorder is specified by the parameter $\gamma = (e^2/\varepsilon
l)/\Gamma$ ($\varepsilon$ is the  dielectric constant).  The solution
of the Hartree-Fock equations is greatly facilitated by the
fact that matrix elements of the exchange potential\cite{mgdm} can be
expressed in terms of the electron density rather than the density
matrix; technical details of these `self-consistent field' calculations
have been explained elsewhere.\cite{aers,lit45}

Localization properties were investigated by evaluating
participation ratios for self-consistent Hartree-Fock
eigenfunctions, $\varphi_\alpha$:
\begin{equation}
P_\alpha \equiv \frac{\big[\int d\vec{r}|\varphi_\alpha
(\vec{r})|^2\big]^2}{L^2\;\int d\vec{r}|\varphi_\alpha (\vec{r})|^4}.
\label{eq:participation}
\end{equation}
Figure~\ref{fig1} displays ${\rm ln}P_\alpha$ as a function of $\nu$
for $\gamma = 0$ and $0.4$ and $\nu_F =1/2$.  (We reserve the symbol
$\nu_F$ for the filling factor at the Fermi level while $\nu$ denotes
the fraction of the density of states below a particular energy.
Most of the results reported here were calculated for the case $\nu_F = 1/2$
for which the extended state occurs at the Fermi energy.)
The participation numbers are shifted slightly upward
by interactions near $\nu = 1/2$ and appear to have a cusp
in their dependence on $\nu$
(Note that no cusp appears in the dependence on energy since the linearly
vanishing density of states implies that $|\nu -1/2 | \propto (E-E_c)^2$
in the presence of interactions.)
The fact that  participation numbers are not greatly changed when the
localization length is smaller than our finite system sizes suggests
that localization lengths are quantitatively unchanged
by interactions.
We infer from this that electron-electron interactions do not affect the
critical properties of localization in any significant way and that
the exponent remains
$7/3$, in agreement with recent experiments\cite{lit80}.

To examine the behavior of the localization length in more detail we
have investigated the system-size dependence of our results.
The left panel of Fig.2 shows the dependence
of ${\rm ln}P_\alpha$ on ${\rm ln}N_\varphi$
in the absence of electron-electron interactions for
$\nu =0.5$, $\nu =0.066 $ and $\nu =0.0022$.
The right panel of Fig.2 shows similair results in
the presence of electron-electron interactions.  Results are
calculated at the filling factors $ 0.5 , 0.081 $ and $0.056$
for $\nu_F = 1/2$ and at $ 0.538, 0.147$ and
0.019 for $\nu_F = 1/5$. For localized states we expect
that $P \propto (\xi/L)^2 \propto \xi^2/N_{\varphi}$.
The results displayed in the left and right panels
show that states in the Landau level
tails are well localized whether the Fermi energy is
at $E_c$ or located in the Landau level tail.  The results shown
at $\nu =$ 0.5 and 0.538 are in a regime where the localization
length exceeds the
system size.  In this regime the participation ratio for non-interacting
electrons is known\cite{aoki} to have a powerlaw dependence on
system size, $ P \propto N_{\varphi}^{-\lambda}$ where the exponent
$\lambda$ is related to the multifractal character\cite{fractal} of the
extended states responsible for the anomolous diffusion\cite{ad}
which occurs when $E_F=E_c$.  The results in Figure~\hbox{\ref{fig2}}
show that the self-consistent Hartree-Fock eigenfunctions in the critical
regime have the same fractal properties as the critical wavefunctions for
non-interacting electrons.
$\lambda$ is given by $ D[2] = 2 (1 - \lambda)$
where $D[2]$ is a measure of the structure of the multifractal.
{}From Fig. \ref{fig2} we estimate that $ D[2] \sim 1.6$ in
agreement with earlier work\cite{aoki,fractal}.

In Figure~\ref{fig3} we show, $\Gamma/\Delta E$, $\delta E/\Gamma$
and $g_T$ as
a function of $\nu$ for $\nu_F = 1/2$, $N_{\varphi}=72$ and
$N_D =1412$ for both interacting and non-interacting
electrons.  Our results for non-interacting electrons are
in agreement with earlier\cite{ando,hanna} work.
As expected, interactions lead to an increase in
$\Delta E $ near $E_F$ corresponding to the density-of-states
suppression.   However, $\delta E$ is also increased and the
net result is that the Thouless number is unchanged to within
the accuracy of our calculations.
In agreement with Hanna {\it et al.} we find that the peak Thouless numbers
appear to already approach their thermodynamic limit for $N_\varphi \sim 50$.
Our numerical results suggest that the peak Thouless numbers
are independent of interactions.  The Thouless
numbers away from $E_F$ are larger for finite-size systems
in the presence of the interactions,
presumably because the screening changes the nature of the disorder
potential.  For bigger system sizes we expect\cite{lit45}
that $\Delta E \propto 1/L$
at $E_F$ which implies that $\delta E$ is also $ \propto 1/L$.
If Eq.(~\ref{eq:sigmaxx}) were exact, the critical conductivity would
be proportional to the peak Thouless number. Our Thouless
numbers, both with and without
interactions, would correspond to peak conductivities
$\approx 0.2 (e^2/h)$. This value is smaller than both the
expected critical conductivity $e^2/2h$ and the
self-consistent Born approximation peak conductivity,
$e^2/h\pi$. To understand these quantitative relationships
the connection between $g_{T}$ and  $\sigma_{xx}$ must be established
under conditions more general that those considered in \onlinecite{lit100}.

In conclusion, we have shown that for the metal-insulator
transition of the integer quantum Hall effect
Thouless numbers defined in terms of the sensitivity of
Hartree-Fock eigenvalues to changes in boundary conditions,
have a critical value which appears to be idependent of the
strength of electron-electron interactions despite large
changes in the density of states at the Fermi energy.
Our work adds further motivation to efforts to
clarify the connection between Thouless numbers and
conductivities, in the presence of interactions and at
a finite magnetic field.

We have benefited from useful suggestions by E. Akkermans, P. Coleridge,
S.M.Girvin, C. Hanna, B. Huckestein, and B. Shklovskii.  This work was
supported by the National Science Foundation under grant
DMR-9416906.

\begin{figure}
\caption{ Participation ratios as a function of $\nu $ for $\gamma = 0$
(crosses) and $\gamma = 0.4$ (squares). The Hartree-Fock results were
obtained for the case of  $\nu_F = 1/2$ and $N_{\varphi} = 200$ by
averaging over $N_D=16$ disorder realizations.
\label{fig1} }
\end{figure}

\begin{figure}
\caption{Left panel: ${\rm ln}P_\alpha$ vs. ${\rm ln}N_\varphi$
for $E = 0,-0.35, -0.6 \Gamma$ in the absence of Coulomb
interactions ($\gamma = 0$).
The filling factors for these energies are $0.5$, $0.066$ and
$0.0022$ respcectively.  The slope of the lines is $\approx -1.0$
for $ E \ne 0$ and $\approx 0.2 $ for $ E = 0$.
Right panel: Same as in the left panel but for $\gamma = 0.4$.
Results were calculated at the filling factors
$ 0.5 , 0.081 $ and $0.056$ for $\nu_F = 1/2$ and at $ 0.538, 0.147$ and
0.019 for $\nu_F = 1/5$. The Hartree-Fock eigenenergies at these filling
factors
are $ -0.225 , -0.6 $ and $-0.8\Gamma$ for $\nu_F = 1/2$ and
$-0.05 , -0.40 $ and $-0.65\Gamma $ for $\nu_F =1/5$.  The
square, circle, and triangle symbols show results for $\nu_F = 1/2$
while the pluse, cross, and diamond symbols show results for $\nu_F = 1/5$.
\label{fig2} }
\end{figure}

\begin{figure}
\caption{$L^{2}D_{\Delta E}(E)/N_{\phi}$, $<\delta E>/\Gamma$
and $0.5g_{T}\pi^{2}$ as a function of $\nu$ for
$\gamma = 0$ (left panels) and $\gamma = 0.4$ (right panels).
$L^{2}D_{\Delta E}(E)$ is the average number of states in the interval
$\Delta E$ around E per disorder realization. Here $N_{D}=1412$
and $\nu_F = 1/2$. \label{fig3}}
\end{figure}
\end{document}